\documentstyle [12pt] {article}
\title {A  GENERAL RELATIVISTIC GENERALIZATION OF BELL INEQUALITY}

\author {Vladan Pankovi\'c\\
Physics Department,Faculty Science \\21000 Novi Sad, Trg Dositeja
Obradovi\'ca 4. \\Serbia and
Montenegro,\\vladanp@gimnazija-indjija.edu.yu}
\date {}
\begin {document}
\maketitle

\vspace {1.5cm}

\begin {abstract}
In this work a general relativistic generalization of Bell
inequality is suggested. Namely,it is proved that practically in
any general relativistic metric there is a generalization of Bell
inequality.It can be satisfied within theories of  local
(subluminal) hidden variables, but it cannot be satisfied in the
general case within standard quantum mechanical formalism or
within theories of nonlocal (superluminal) hidden variables. It is
shown too that within theories of nonlocal hidden variables but
not in the standard quantum mechanical formalism a paradox appears
in the  situation when one of the correlated subsystems arrives at
a Schwarzschild black hole. Namely, there is no way that black
hole horizon  obstructs superluminal influences between spin of
the subsystem without horizon and spin of the subsystem within
horizon,or simply speaking,there is none black hole horizon nor
"no hair" theorem for subsystems with correlated spins. It implies
that standard quantum mechanical formalism yields unique
consistent and complete description of the quantum mechanical
phenomenons.
\end {abstract}
\vspace {1.5cm}

 \vspace {1.5cm}

\newpage

There are many attempts [1]-[5] of the special relativistic
generalization of the remarkable Bell inequality [6].They are
based, roughly speaking, on the different attempts of the
specially relativistically generalized Bell states and observable.

In this work a general relativistic generalization of Bell
inequality  will be suggested. It will be based on a simple
approximate, called hybrid, description of the quantum supersystem
and its distant subsystems, where, roughly speaking, "spin parts"
of these systems are described effectively quantum mechanically
without any general relativistic corrections, while "particle
parts" of these systems are described effectively by general
theory of relativity without any quantum mechanical corrections  .
In this way not only problems of the general relativistic
generalization of quantum mechanical dynamics but also problems of
the general relativistic generalizations of usual measurement
procedures will be effectively removed.

It will be proved that then in practically any general
relativistic metric of the space-time  there is a generalization
of Bell inequality. Such generalized Bell inequality can be
satisfied within theories of the  local hidden variables that,
roughly speaking, hold subluminal dynamics and suppose that
Hilbert space of the quantum states represents a formal, even
artificial construction over usual space-time as the basic
physical space. But generalized Bell inequality cannot be
satisfied in the general case within standard quantum mechanical
formalism [7]-[9]. This formalism, roughly speaking, supposes that
only Hilbert space (eg. : so-called orbital Hilbert space with
coordinate and momentum observable and theirs analytical
functions; spin Hilbert space with spin observable and theirs
analytical functions; tensorial product of the orbital and spin
Hilbert spaces; etc.) represents basic physical space while usual
space-time represents an approximation of an especial, so-called
orbital Hilbert space. Generalized Bell inequality cannot be
satisfied too within theories of the nonlocal hidden variables
that, roughly speaking, hold superluminal dynamics and suppose
that Hilbert space of the quantum states represents a formal, even
artificial construction over usual space-time as the basic
physical space. Meanwhile, it will be shown that within theories
of the nonlocal hidden variables but not within standard quantum
mechanical formalism a paradox appears in the  situation when one
of the correlated subsystems arrives at a Schwarzschild black
hole. Namely, there is no way that black hole horizon obstructs
superluminal influences between spin of the subsystem without
horizon and spin of the subsystem within horizon, or simply
speaking, there is none black hole horizon for subsystems with
correlated spins. It implies that standard quantum mechanical
formalism, including its usual, Copenhagen interpretation [10],
[11], yields unique consistent and complete description of the
quantum mechanical phenomenons.

So, let after decay of an initial nonstable quantum system with
total zero spin a new quantum supersystem, $1+2$, that holds two
quantum subsystems, $1$ and $2$, originates. Let this $1+2$ be
described by following quantum state
\begin {equation}
|\Psi^{1+2}>=|S^{1+2}>|\Psi^{1}(x^{1\mu})>|\Psi^{2}(x^{2\mu})>
\end {equation}
Let in (1)
\begin {equation}
|S^{1+2}>= (\frac {1}{2})^{-\frac
{1}{2}}[|+1^{1}>|-1^{2}>-|-1^{1}>|+1^{2}>]
\end {equation}
be a correlated quantum state of the spin of $1+2$ where $|j^{{\it
i}}>$ represents $j$ quantum state of the z-component of  spin
observable $(\frac {\hbar}{2} \hbar {\sigma}_{z}$ that acts over
spin Hilbert space of subsystem ${\it i}$ for $j= -1,+1$ and ${\it
i} =1,2$. This correlated state belongs, as it is well known, to
spin Hilbert superspace of supersystem $1+2$ that represents the
tensorial product of the spin Hilbert spaces of the subsystems.
Also, let in (1) $|\Psi^{\it i}(x^{{\it i}\eta})>$ be the quantum
state (that depends of the space-time coordinates $x^{{\it
i}\eta}$ ) from orbital Hilbert space of the subsystem ${\it i}$
for ${\it i}=1,2$ and $\eta=0,1,2,3$. According to standard
quantum mechanical formalism orbital and spin Hilbert space of any
subsystem are principally ${\it different}$ and ${\it
independent}$. It means that  $|\Psi^{1+2}>$ represents quantum
state that belong to a complete Hilbert space of $1+2$ that
represents tensorial product of all subsystemic orbital and spin
Hilbert spaces.

It will be supposed that $|\Psi^{{\it i}}(x^{{\it i}\eta})>$
represents a slowly dissipated wave packet so that it can be
effectively, i.e. in a satisfactory approximation, represented by
a particle with four-coordinate $<x^{{\it i}\eta}>$ and
momentum-energy $<p^{{\it i}\eta}>$ (where $<x^{{\it i}\eta}>$ and
$<p^{{\it i}\eta}>$ represent corresponding coordinate and
momentum-energy observable) for $\eta=0,1,2,3$  and   ${\it
i}=1,2$. Also, it will be supposed that these wave packets
propagate in the opposite directions ($1$ in the "left" and $2$ in
the "right" in respect to some initial point ${\it o}$) along some
curved line in the three dimensional space with $x^{j}$
coordinates for $j=1,2,3$  while in the complete space-time with
$x^{\mu}$ coordinates and with metric tensor $g_{\mu \nu}$ for
$\mu \nu=0,1,2,3$ determined by general relativistic dynamics
equations these two wave packets propagate along two different
geodesic lines with common initial point $O$.

In this way a satisfactory approximation of the description of
$1+2$ is done. Obviously, within this "hybrid" approximation "spin
part" of $1+2$, i.e. "spin parts" of $1$ and $2$, are described
exactly quantum mechanically by $|S^{1+2}>$, while
"four-coordinate part" of $1+2$, i.e. "particle parts" of $1$ and
$2$, precisely theirs propagations through space-time are
described effectively, i.e. in a satisfactory approximation by
general theory of relativity.

Denote by $L$ a point on the geodesic line of $1$ in which a local
Lorentzian referential frame $S_{L}$ is defined. Within $S_{L}$ a
spin observable on $1$ can be measured in the usual, well-known
way in a direction with unit norm  vector $a^{\alpha}_{L}$.

Denote by $R$ a point on the geodesic line of $2$ in which a local
Lorentzian referential frame $S_{R}$ is defined. Within $S_{R}$ a
spin observable on $2$ can be measured in the usual, well-known
way in a direction with unit norm vector $b^{\beta}_{R}$.

As it is well-known within general Riemanian geometry arbitrary
vector cannot be parallel transferred along a geodesic line.
Meanwhile we can realize a transfer of $b^{\beta}_{R}$ along given
geodesic line of $2$ from $R$ in $O$ and further along geodesic
line of $1$ from $O$ in $L$ so that these transfer retriers
minimally from corresponding parallel transfers. In the general
case, obtained by given transfers, new vector does not represent
an unit norm vector in $S_{L}$. To be a unit norm vector in
$S_{L}$ given transferred vector must be firstly normalized on the
one and then projected orthogonally, if it is possible, or in a
way that minimally retries from orthogonal projection, in $S_{L}$.
This projection we shall denoted by  $w(b^{\beta}_{R})
b^{\beta}_{RL}$ where $b^{\beta}_{RL}$ represents corresponding
unit norm vector in $S_{L}$  while $w(b^{\beta}_{R})$ represents
absolute value of given projection that belongs to $[0,1]$
interval. It is not hard to conclude that concrete form of the
obtained projection depends of the concrete form of both geodesic
lines, the metrical tensor $g_{\mu \nu}$ and $b^{\beta}_{R}$ which
will not be considered with details.

Introduce following expression
\begin {equation}
P(a^{\alpha}_{L},b^{\beta}_{R}) = -g^{0}_{L \alpha \beta}
a^{\alpha}_{L} b^{\beta}_{RL}w^{2}( b^{\beta}_{R})
\end {equation}
or, shortly,
\begin {equation}
P(a^{\alpha}_{L},b^{\beta}_{R}) =  - a_{L}b_{LR} w^{2}(b_{R})
\end {equation}
Here $ g^{0}_{L \alpha \beta} a^{\alpha}_{L} b^{\beta}_{RL}$
represents scalar product of  $ a^{\alpha}_{L}$ and $
b^{\beta}_{RL}$ in $S_{L}$ with $ g^{0}_{L \alpha \beta}$ that
represents metrical tensor of the local Lorentzian metric in $L$.
We shall consider that this expression (3) represents a general
relativistic generalization of the quantum mechanical expression
for average value of the product of the spin along $a^{alpha}_{L}$
on $1$ and spin along $ b^{\beta}_{R}$ on $2$ from aspect of
$S_{L}$.

It can be pointed out that a general relativistic generalization
of the quantum mechanical expression for average value of the
product of the spins on $1$ and $2$ from aspect of $S_{R}$R can be
realized in an analogous way which will not be considered.

Define following functions of a general relativistic scalar
$\lambda$
\begin {equation}
A(a^{alpha}_{L},\lambda) = \pm 1  \hspace{1cm}  for \hspace{1cm}
\forall \lambda, a^{alpha}_{L}
\end {equation}
or shortly
\begin {equation}
A(a_{L},\lambda) = \pm 1   \hspace{1cm}  for \hspace{1cm} \forall
\lambda , a_{L}
\end {equation}
and
\begin {equation}
B(b^{\beta}_{RL},\lambda) = \pm w^{2}(b^{\beta}_{R})  \hspace{1cm}
for \hspace{1cm} \forall \lambda,b^{\beta}_{RL}
\end {equation}
or shortly
\begin {equation}
B(b_{RL},\lambda) = \pm w^{2}(b_{R})    \hspace{1cm}   for
\hspace{1cm} \forall \lambda, b_{RL}
\end {equation}
We shall consider that $\lambda$ represents a hidden variable  so
that  expression (5), i.e. (6)  determines spin value on $1$ and
expression (7), i.e. (8) - spin value on $2$, more precisely than
standard quantum mechanical formalism.

Suppose that there are following correlations
\begin {equation}
A(a_{RL},\lambda)= - w^{-2}(a_{R})B(a_{RL},\lambda) \hspace{1cm}
for \hspace{1cm} \forall \lambda, a_{R}
\end {equation}
i.e.
\begin {equation}
B(a_{RL},\lambda) = - w^{2}(a_{R}) A(a_{RL},\lambda)  \hspace{1cm}
for \hspace{1cm} \forall \lambda, a_{R}
\end {equation}

Finally, suppose that following is satisfied
\begin {equation}
P(a_{L},b_{RL})=\int \rho  (\lambda) A(a_{L},\lambda)
B(b_{RL},\lambda) d\lambda
\end {equation}
where $\rho (\lambda)$ represents hidden variables probability
density supposed as a general relativistic scalar. This expression
can be considered as the probabilistic reproduction of the general
relativistic and quantum mechanical average value of the spins
product $P(a_{L}, b_{RL})$ by means of the hidden variables. It
can be added that since given average value of  the spins product
is determined by probabilistic distribution of hidden variables
common for $1$ and $2$ these hidden variables have a local, i.e.
subluminal character.

According to (10) expression (11) turns in
\begin {equation}
P(a_{L},b_{RL})= - \int \rho (\lambda) w^{2}(a_{R})
A(a_{L},\lambda) A(b_{RL},\lambda) d\lambda
\end {equation}

Then,it follows
\[P(a_{L},b_{RL})-P(a_{L},c_{RL}) =\]
\begin{equation}
= -\int\rho(\lambda)[w^{2}(b_{R}) A(a_{L},\lambda)
A(b_{RL},\lambda) - w^{2}(c_{R}) A(a_{L},\lambda)
A(c_{RL},\lambda) ]d\lambda =
\end{equation}
\[ = -\int \rho (\lambda) [w^{2}(b_{R}) - w^{2}(c_{R})
A(a_{L},\lambda) A(c_{RL},\lambda) ] A(a_{L},\lambda)
A(b_{RL},\lambda) d\lambda \]

where $c_{R}$ is some other unit vector in $S_{R}$. Further, from
(13) it follows
\begin {equation}
|P(a_{L},b_{RL})-P(a_{L},c_{RL})| \leq
w^{2}(b_{R})+P(b_{RL},c_{RL})
\end {equation}
which represents {\it a general relativistic generalization} of
Bell inequality.

Introduction of (4) in (14) yields
\begin {equation}
|-w^{2}(b_{R})a_{L}b_{RL} + w^{2}(cR)a_{L}c_{RL}| \leq
w^{2}(b_{R}) - w^{2}(c_{R})b_{RL}c_{RL}
\end {equation}
or, since $a_{L}$, $b_{RL}$ and $c_{RL}$RL represents unit norm
vectors in $S_{L}$,

\[|- a_{L}(w^{2}(b_{R})b_{RL} - w^{2}(c_{R})c_{RL}) | \leq
w^{2}(b_{R}) b_{RL}b_{RL} - w^{2}(c_{R})b_{RL}c_{RL}= \]
\begin {equation}
= b_{RL}(w^{2}(b_{R})b_{RL} - w^{2}(c_{R})c_{RL})
\end {equation}
and
\begin {equation}
| w^{2}(b_{R})b_{RL} - w^{2}(c_{R})c_{RL}| | \cos \varphi| \leq |
w^{2}(b_{R})b_{RL} - w^{2}(c_{R})c_{RL}| \cos \theta
\end {equation}
where $\varphi$ represents the angle between $a$ and $
w^{2}(b_{R})b_{RL} - w^{2}(c_{R})c_{RL}$ while $\theta$ represents
the angle between $b_{RL}$ and $ w^{2}(b_{R})b_{RL} -
w^{2}(c_{R})c_{RL}$

From (17) it follows
\begin {equation}
|\cos \varphi | \leq \cos \theta     for  | w^{2}(b_{R})b_{RL} -
w^{2}(c_{R})c_{RL}| \neq 0
\end {equation}

But (18) cannot be always satisfied. Namely, for given $b_{RL}$
and $c_{RL}$ angle $\theta$ is determined practically
unambiguously (till $2k\pi$ for $k=1,2,…$ ), while angle $\varphi$
can be different for different $a_{L}$. For this reason, for given
$b_{RL}$  and $c_{RL}$, $a_{L}$, i.e. $\varphi$ can be chosen in
such way that (18) cannot be satisfied.

All this points clearly  that within suggested general
relativistic generalization of the Bell inequality any local
(subluminal) hidden variables theory contradicts to standard
quantum mechanical formalism. Nevertheless, some nonlocal
(superluminal) hidden variables theories can exist that break
suggested generalization of Bell inequality and that are
consistent with standard quantum mechanical formalism.

Suppose that future experiments will show that suggested
generalization of the Bell inequality is broken, i.e. that
suggested hybrid description of $1+2$ by quantum mechanics and
general theory of relativity is correct in supposed limits. It
would mean too that some nonlocal (superluminal) hidden variables
theories can exist. But we shall show now that within such
nonlocal hidden variables theories but not within standard quantum
mechanical formalism a paradox appears in the  situation when one
of the correlated subsystems, eg.  $2$ arrives at a Schwarzschild
black hole (without electrical charge and angular momentum).

Namely, in this situation, roughly speaking,
$|\Psi^{1}(x^{1\mu})>$ that belongs to {\it orbital} Hilbert space
of 1 describes "particle part" of 1 without black hole, while
$|\Psi^{2}(x^{2\mu})>$ that belongs to {\it orbital} Hilbert space
of $2$ describes "particle part" of $2$ within black hole. It
means that from general relativistic view point, without quantum
mechanical corrections, "particle parts" of $1$ and $2$ in the
usual, i.e. general relativistic space-time, as well as any two
usual measurement procedures (that considers classical measurement
devices [10], [11]) nearly given "particle parts" respectively,
are {\it absolutely separated} in the space-time by black hole
horizon on the one side.

But, on the other side, according to suppositions, given black
hole is Schwarzschild, i.e. without electrical charge and angular
momentum. It means that it can subluminally gravitationally
dynamically to act only at space-time variables of 1,2 or 1+2, or,
more precisely, only at states of $1$, $2$, or $1+2$ from
corresponding {\it orbital} Hilbert spaces. In other words, from
aspect of the standard quantum mechanical formalism, given black
hole does not any subluminal dynamical influence on $|S^{1+2}>$
that belongs to {\it spin} Hilbert superspace of 1+2. For this
reason $|S^{1+2}>$  stands a correlated quantum state, i.e. a
supersystemic superposition (2) {\it without any separation} of
the "spin part" of supersystem $1+2$ in its subsystems, "spin
part" of $1$ and "spin part" of  $2$. Or, simply speaking, for
"spin parts" of $1$, $2$ and $1+2$ there is none black hole
horizon.

It can be added that, according to suppositions, spin on the
subsystem {\it i} must be measured by the usual measurement
procedure nearly "particle part" of this subsystem, for ${\it i}
=1,2$. Also, as it has been noted, "particle parts" of $1$ and $2$
and corresponding usual measurement procedures are absolutely
separated by black hole horizon. It causes that there is none
possibility for a (sub)luminal, i.e. local communication between
measurement procedures of the "spin part" of $1$ and "spin part"
of  $2$. It is principally similar to situation which we have by
real experimental tests of Bell inequality [12], [13], during
small intervals when measurement devices are distant, i.e.
separated by a space interval. For this reason it is principally
admitable, from standard quantum mechanical formalism view point,
to suppose that spins correlation between of 1 and 2 described by
$|S^{1+2}>$ (2) really exists even if "particle part" of one of
the subsystems is within and other without horizon so that this
correlation cannot be effectively experimentally checked.

For the aspect of the nonlocal hidden variables theories the same
situation must be interpreted in following way. First of all, as
it has been pointed out, within such theories any Hilbert space
represents only a formal, i.e. artificial  but not real physical
space while general relativistic  space-time represents real
physical space. It implies that "spin parts" of subsystems 1 and 2
must be always space-time localized by "particle parts" of 1 and
2. It means that whole subsystem 1 is without black hole while
whole subsystem 2 is within black hole and that these two
subsystems, according to general theory of relativity (without
quantum mechanical corrections) must be absolutely separated by
horizon. Also within nonlocal hidden variables theories the
quantum mechanical spin correlation of the "spin parts" of $1$ and
$2$ represents formally a real superluminal influences between $1$
and $2$. But in this case black hole horizon cannot to destroy
given superluminal influence between $1$ and $2$ in any way.
Simply speaking, for superluminal hidden variables there is
practically none black hole horizon or, as it is not hard to see,
for nonlocal hidden variables theories the important "no hair"
theorem cannot be satisfied (since hidden variables conserves
individuality of any quantum system that arrives in black hole).
It seems very implausible or paradoxical.

All this implies a conclusion that  standard quantum mechanical
formalism, including its usual, Copenhagen interpretation, yields
unique consistent and complete description of the quantum
mechanical phenomenons.

\section {References}

\begin {itemize}
\item [[1]] M.Czachor,Phys.Rev. A {\bf 55},(1997.) ,72.
\item  [[2]] A.Peres,P.F.Seudo,D.R.Terno,Phys.Rev.Lett,{\bf 88},(202.),230402
\item [[3]] A.Peres,D.R.Terno,Rev.Mod.Phys.,{\bf 76},(2004.),93.
\item [[4]] D.Aun,H.-J.Lee,H.Moon,S.W.Hwang,Phys.Rev.A {\bf 67},(2003.),012103
\item  [[5]] W.T.Kim,E.J.Sor,quant-ph/0908127 v2 28 Dec 2004.
\item [[6]] J.S.Bell,Physics,{\bf 1},(1964.),195.
\item [[7]] J.von Neumann,{\it Mathematische Grundlagen der Quanten Mechanik }(Springer Verlag,Berlin,1932.)
\item [[8]] P.A.M.Dirac,{\ it Principles of Quantum Mechanics }(Clarendon Press,Oxford,1958.)
\item [[9]] B.d'Espagnat,{\it Conceptual Foundations of Quantum Mechanics }(Benjamin,New York,1976.)
\item [[10]] N.Bohr,Phys.Rev.,{\bf 48},(1935.),696.
\item [[11]] N.Bohr,{\it Atomic Physics and Human Knowledge},(John Wiley,New York,1958.)
\item [[12]] A.Aspect,P.Grangier,G.Roger,Phys.Rev.Lett.,{\bf 47},(1981.),460.
\item [[13]] A.Aspect,J.Dalibard,G.Roger,Phys.rev.Lett.,{\bf 49},(1982.),1804.

\end {itemize}
\end {document}